# Single frame wide-field Nanoscopy based on Ghost Imaging via Sparsity Constraints (GISC Nanoscopy)

**WENWEN LI[1, 2], ZHISHEN TONG[3,4], KANG XIAO[1, 2], ZHENTAO LIU[3], QI GAO[2], JING SUN[2], SHUPENG LIU[1], SHENSHENG HAN[3*] AND ZHONGYANG WANG[2*]**

[1] Key Laboratory of Specialty Fiber Optics and Optical Access Networks, Shanghai Institute for Advanced Communication & Data Science, Shanghai University, 200444 Shanghai
[2] Shanghai Advanced Research Institute, Chinese Academy of Sciences, Shanghai 201210, China
[3] Key Laboratory for Quantum Optics and Center for Cold Atom Physics of CAS, Shanghai Institute of Optics and Fine Mechanics, Chinese Academy of Sciences, Shanghai 201800, China
[4] Center of Materials Science and Optoelectronics Engineering, University of Chinese Academy of Sciences, Beijing100049, China
*Corresponding author: wangzy@sari.ac.cn

**Abstract**

The applications of present nanoscopy techniques for live cell imaging are limited by the long sampling time and low emitter density. Here we developed a new single frame wide-field nanoscopy based on ghost imaging via sparsity constraints (GISC Nanoscopy), in which a spatial random phase modulator is applied in a wide-field microscopy to achieve random measurement for fluorescence signals. This new method can effectively utilize the sparsity of fluorescence emitters to dramatically enhance the imaging resolution to 80 nm by compressive sensing (CS) reconstruction for one raw image. The ultra-high emitter density of 143 $\mu m^{-2}$ has been achieved while the precision of single-molecule localization below 25 nm has been maintained. Thereby working with high-density of photo-switchable fluorophores GISC nanoscopy can reduce orders of magnitude sampling frames compared with previous single-molecule localization based super-resolution imaging methods.

## 1. INTRODUCTION

The fluorescence microscope is an essential tool for the study of biological processes and biological phenomena. However, features at sizes less than the half of a light wavelength (200-300 nm) cannot be resolved in conventional far-field microscopy because of the optical diffraction barrier. In recent decades, various new techniques aiming at breaking the diffraction barrier have sprung up [1], which have been achieved in two ways: One spatially modulate the target with patterned illumination, such as stimulated emission depletion microscopy (STED) [2], reversible saturable optical fluorescence transitions (RESOLFT) [3] and structured illumination microscopy (SIM) [4]. These techniques require point scanning or complex illumination patterns to construct one high-resolution image, which restrict the speed of live cell imaging. The other use stochastic single-molecule switching and fluctuation of fluorophores, such as photo-activated localization microscopy (PALM) [5], stochastic optical reconstruction microscopy (STORM) [6], and point accumulation for imaging in nanoscale topography (DNA-PAINT) [7], super-resolution optical fluctuation imaging (SOFI) [8], etc. Although by wide-field imaging rather than sequentially point scanning, these techniques need collect a large number of raw images (thousands to tens of thousands) to trace fluctuation/blinking information of fluorophores, and ensure that no close-by emitters can be switched on simultaneously, which largely restrict the density of emissive fluorophores per frame [9]. Therefore their applicability for live cell imaging is significantly confined by the low temporal resolution owing to the long sampling time and low emitter density.

To overcome the limit of temporal resolution in live-cell imaging, several super-resolution methods have been developed. For scanning-based techniques, STED and RESOLFT can image with very high temporal resolution by reducing the field of view [10]. For wide field imaging, live SIM uses 9 frames with 3.7 to 11 Hz frame rates to construct one high-resolution output image at 100 nm resolution [11], the temporal resolution of STORM can be improved either by shortening sampling time using fast-switching dyes [12] and fast scientific-CMOS cameras [13], or by increasing the emitter density, such as DAOSTORM by fitting clusters of overlapped spots (about 3-4 $\mu m^{-2}$) [14], CS-STORM based on compressed sensing algorithm (8.8 $\mu m^{-2}$) [15], Deep-STORM harnessing Deep Learning method (6 $\mu m^{-2}$) [16]. Other methods to handle a high density of overlapping emitters are SPIDER (15 $\mu m^{-2}$) [17], SPARCOM (35 $\mu m^{-2}$) [18] and dictionary learning [19] using different sparsity representation in the transformation domain for original signal to obtain sparse information. These methods can reduce the acquisition time down to hundreds to thousands of frames, however worsen the spatial resolution (30-60 nm) and rely on complex data-processing. There is still a big challenge for a live cell imaging technique that can combine spatial super-resolution with high-speed imaging in large fields of view.

Here we develop a wide-field nanoscopy based on ghost imaging via sparsity constraints (GISC Nanoscopy) that is capable of 80 nm resolution for single frame imaging. The technique relies on the naturally sparse fluorescence signals and combines ghost imaging (GI) method and compressive sensing (CS) theory. CS as a signal processing method utilizes the sparsity property of the signal to achieve super-resolution imaging with random measurement [20-22]. GI, which is based on the quantum or classical correlation of fluctuating light fields, can non-locally image an unknown object by transmitting pairs of photons through a test and a reference paths without point scanning [23-26].GI method of extracting the imaging information is based on "global random" measurement and the fluctuating light filed obey Gaussian statistical distribution, which satisfies the restricted isometry property (RIP) required for CS [20,21]. Therefore, combining GI and CS, a way of combining optical imaging technique with modern information theory, GISC has made many potential applications including super-resolution imaging [27-29], GISC spectral camera [30], GISC lidar [31], and X-ray diffraction imaging [32].

GISC nanoscopy applies a spatial random phase modulator in conventional fluorescence microscope to form random speckle patterns and code the detected fluorescence signals. A random measurement matrix composed of speckles corresponding to different positions of the sample plane is built, which transform the measurement matrix of signal to satisfy the CS requirement, therefore the technique can effectively reconstruct a super-resolution imaging via sparsity constraints from one low-resolution wide-field speckle image. We demonstrate reconstructions in simulation and experiment with the considerably improved spatial resolution, reducing the number of required exposures to a single-frame exposure, and expedite the fluorescence imaging process. The technique can apply for any fluorescent specimen without complex illumination modes or the intrinsic blinking/fluctuation mechanism of fluorescent molecules. Meanwhile we also achieve ultra-high density (up to 143 μm$^{-2}$) of single-molecule localization with intensity and position information, and the localization precision is still below 25 nm. Therefore we can combine with single-molecule localization-based super-resolution techniques (GISC-STORM) to have orders-of-magnitude shorter sampling times than the previous approaches without a worse spatial resolution in return. Moreover, the main factors influencing the quality of super-resolution imaging of signal-to-noise ratio and contrast are investigated.

## 2. METHODS

### 2.1 Theory

According to the theory of GISC imaging which is applied to GISC camera [30], the fluorescent light field is modulated into speckle pattern by a random phase modulator mounted before detector. GISC nanoscopy consists of two parts: 1) Calibration process, we collect the speckle patterns $I^r_{(i',j')}(i,j)$ corresponding to each position in the sample plane as the priori information, which can be measured on-line or pre-determined. 2) Imaging process, we can obtain the speckle pattern intensity distribution $I^t_{(i',j')}$ from all fluorophores within the imaged specimen to achieve a super-resolution reconstructed image by calculating the second-order intensity correlation between the calibration speckles and one imaging speckle. The second-order correlation function is expressed as [26]

$$\Delta G^{(2)}(i,j) = \langle I_r(x_r)I_t(x_t) \rangle \approx \frac{1}{M_{i'}M_{j'}} \sum_{i'=1}^{M_{i'}} \sum_{j'=1}^{M_{j'}} I^r_{(i',j')}(i,j) I^t_{i'\,j'} \qquad (1)$$

Where $I_r(x_r)$ and $I_t(x_t)$ respectively denote the intensity distribution of speckle patterns in the reference and test path, $x_r$ and $x_t$ represent the horizontal coordinate of the detection plane. $I^r_{(i',j')}(i,j)$ is the speckle intensity at pixel $(i',j')$ recorded by the detector in calibration process, corresponding to the fluorescence signal at grid $(i,j)$ in the sample plane. $I^t_{(i',j')}$ denotes the speckle intensity of the imaged specimen in imaging process. $i = 1, 2, ...., N_i$ and $j = 1, 2, ..., N_j$ represent the horizontal and vertical grid coordinates in the sample plane respectively, $i' = 1, 2, ..., M_{i'}$ and $j' = 1, 2, ..., M_{j'}$ respectively represent the horizontal and vertical pixel coordinates in the detection plane.

In GISC nanoscopy, the resolution of imaging is determined by three aspects: 1) Resolution of GI, which is related to the mutual-correlation function of speckle patterns and can be optimized by adjusting the position of random phase modulator between the tube lens and detector and the parameters of the random phase modulator [30]. **Fig. 2a** present the mutual-correlation function of the speckle and point spread function (PSF) of fluorescence signal measured in experiment, the resolution of GI can be improved by a factor of $\sqrt{2}$ compared to conventional wide-field microscope, which is the same as the resolution of confocal microscope reached. The capacity of GI resolution has been demonstrated theoretically based on the second-order correlation of light fields [33,34] and can be further improved by calculating the high-order correlation. 2) Sparse constraint, which is utilized as an image processing method to recover high frequency content lost in the measurement process. The resolution of imaging increased by 2 times based on sparsity constraint has been demonstrated mathematically by Candès [22,29]. 3) Shifted with subpixel precision, which allows super-resolution reconstruction from observed multiple low-resolution measurements by the reconstruction approaches [35]. Therefore the higher resolution of GISC nanoscopy can be achieved by utilizing the difference of speckle centers from its subpixel precision random movement in each sampling.

CS theory has proved that imaging information with sparsity can be accurately reconstructed when two key conditions are satisfied: the imaging object's sparsity in the representation basis and random measurement. The "global random" measurement of GI and the fluctuating light filed obeying Gaussian statistical distribution satisfies strictly the restricted isometry property (RIP) required for CS. Therefore, we re-formulate [Eq. (1)] in the CS framework to realize single frame super-resolution imaging. The original signal intensity $I(i, j)$ is denoted as a column vector $X\left(N \times 1, N = N_i \times N_j\right)$. The speckle intensity $I'_{(i',j')}$ detected in imaging process is arranged as a column vector $Y\left(M \times 1, M = M_{i'} \times M_{j'}\right)$. After $N$ calibration measurements, we can build a measurement matrix $A\left(M \times N\right)$, each of the calibrated speckle intensity $I'_{(i',j')}(i, j)$ is reshaped as one column vector of the matrix $A$. CS theory is expressed as [20]

$$\min \|x\|_0 \qquad s.t. \ Y = AX \qquad (2)$$

Where $Y$ is the observation of a set of "global random measurements", $A$ is the random measurement matrix, and $X$ is the original image.

We utilize two common compressed sensing algorithms to solve [Eq. (2)]: Gradient Projection for Sparse Reconstruction Method (GPSR) belonging to Solving Convex Optimization Equations [36] and Orthogonal Matching Pursuit Method (OMP) belonging to Greedy Iterative Algorithms [37]. When using GPSR algorithm [Eq. (2)] can be transformed to implement the L1 norm minimization: $\min \|x\|_1 \, s.t. \|Ax - y\|_2 \leq \varepsilon$ ( $\varepsilon$ is related to noise), then simplified to a quadratic programming. GPSR has advantages in noise reduction and high reconstruction accuracy. OMP algorithm solving [Eq. (2)] is to find the best matching value of $X$ by iteration. It has faster calculation speed comparing to other CS algorithms, which is more suitable for this technique due to the large amount of data for wide field imaging. However, OMP algorithm has a slightly worse recovery accurate than GPSR.

## 2.2 Experimental setup and calibration process

The experimental setup and imaging process of GISC nanoscope are shown in **Fig. 1**. We performed the experiment on a conventional inverted microscope (IX83, Olympus, USA) with a total internal-reflection excitation scheme (cellTIRF-4Line, Olympus) and an oil-immersion objective (Olympus, 100X, Numerical aperture: 1.49). A spatial random phase modulator (DG10-1500-A, Thorlabs) is mounted after the focal plane of imaging system to generate random speckle patterns of fluorescence signal, and subsequently a low magnification objective (UPlanSApo 10X, Olympus) is used to magnify the speckles which are detected as the matrix $Y$ using a scientific-CMOS (Prime 95B, Photometrics, a pixel size corresponds to 50 nm in the sample plane). Fluorescence signal is directly detected by an EMCCD (Evolve 512, Photometrics, a pixel size corresponds to 160 nm in the sample plane). 532 nm (OBIS-532 nm-LS-80 mW, Coherent) and 640 nm (OBIS-640 nm-LX-75 mW, Coherent) lasers are optional as different excitation source, which were filtered by a band-pass excitation filter (ZET532/640x, Chroma) and reflected by a multi-band dichroic (ZT532/640rpc, Chroma) onto the sample. And the collected fluorescence from the sample was filtered by a band-pass emission filter (ZET532/640m-TRF, Chroma). In order to shorten the calibration and reconstruction time, we use the block calibration and reconstruction method. The sample plane is divided into multiple blocks and we analyze and optimize these blocks separately, the size of each block is 1.72 μm×1.72 μm. Each block is subdivided into 86×86 grids as the matrix $X$ (20 nm per grid) to ensure imaging accuracy (According to Nyquist criterion). We take one block as an example in the simulations and experiments.

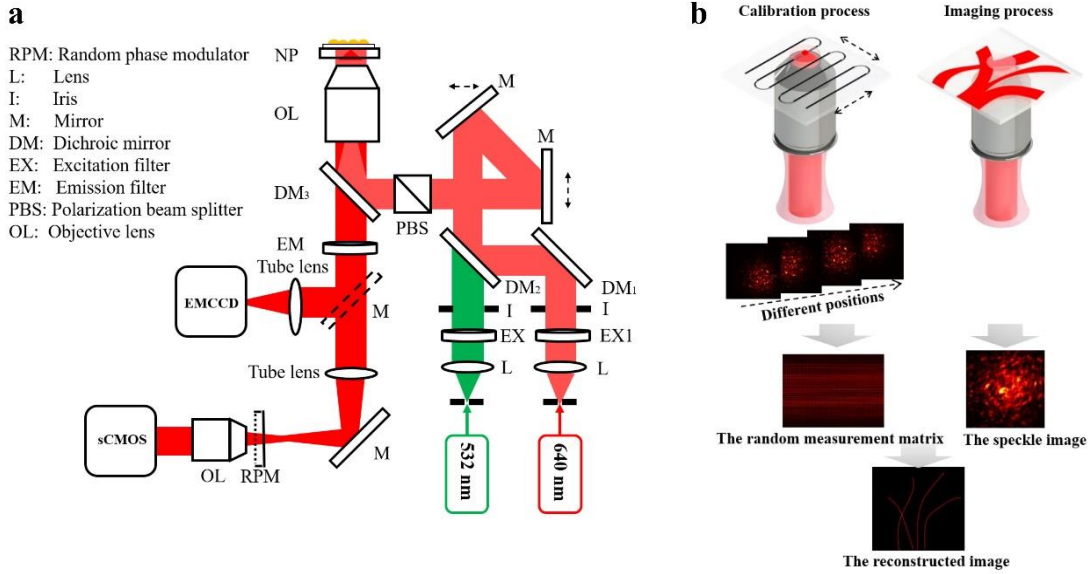

**Fig. 1.** Schematic diagram of experimental setup and imaging. (**a**) Experimental setup, a random phase modulator with a low magnification objective (10X) is set before sCMOS in a conventional inverted microscope, to form speckle patterns of fluorescence signal. Fluorescence image is directly recorded by EMCCD. (**b**) Calibration process and imaging process, all speckle patterns generated from each position of the sample plane are recorded as the random measurement matrix in calibration process and one speckle image from the actual imaged sample are obtained in imaging process, then a super-resolution image can be reconstructed by compressed sensing.

Single microsphere (20 nm) in the calibration area is selected as the calibration source, which is mounted on a Nanopositioning stage (E725, PI). A series of speckle images of microsphere at different positions (spacing 100 nm) are obtained accurately by controlling stage and sCMOS synchronously. Then the normalized measurement matrix $A$ is built with a pixel spacing of 20 nm (matches the grid size of the matrix $X$ ) by bilinear interpolation from a series of speckle images. To ensure the accuracy of the reconstruction, two key issues should be addressed in calibration process: 1) Selection of calibration source, which requires high fluorescence brightness, high stability and small difference in size and spectral with the dye used in imaging process. Therefore we use microspheres (20 nm) as the calibration source because the size difference with fluorophore is negligible to the effects of noise and pixel size of detector demonstrated by system simulation or the deconvolution, and the spectral difference is limited by the bandpass filter selected according to the spectral resolution of GISC imaging [30]. Thus any fluorescent sample can be used at the same excitation wavelength with the microsphere, which makes this technique more widely used in sample preparation and dye selection. 2) Translation invariance, in the calibration process, the accuracy of the translation and the stability of the light source need to be ensured. Therefore the mutual-correlation between the two adjacent speckle images is above 0.95 to reduce the interpolation error. Axial drift need be corrected by a device (IX3-ZDC2, Olympus) and lateral drift for the reconstructed image is calibrated according to the drift curve of system obtained before the calibration process.

## 3. RESULT

### 3.1 Simulation with the real speckle patterns

We use the real speckle patterns obtained in calibration process of experiment to analyze the resolution of this technique and its influencing factors. In order to quantify the improved resolution of our technique, we generated rings with different spacing (80 nm, 120 nm, 240 nm) composed of high density molecules. The number of photons emitted from a molecule follows a log-normal distribution (the peak of 4 000 photons with a standard deviation of 1 700 photons), which matches the experimentally measured single-molecule photon distribution of Alexa Fluor 647. Then the photon number need multiply the peak attenuation coefficient due to the photon diffusion when passing through the phase modulator. As observed In **Fig. 2b**, the single frame reconstruction results with real speckle patterns show that the spatial resolution of GISC nanoscopy is as high as 80 nm according to the Rayleigh criterion, which has a significant improvement compared to corresponding wide field imaging.

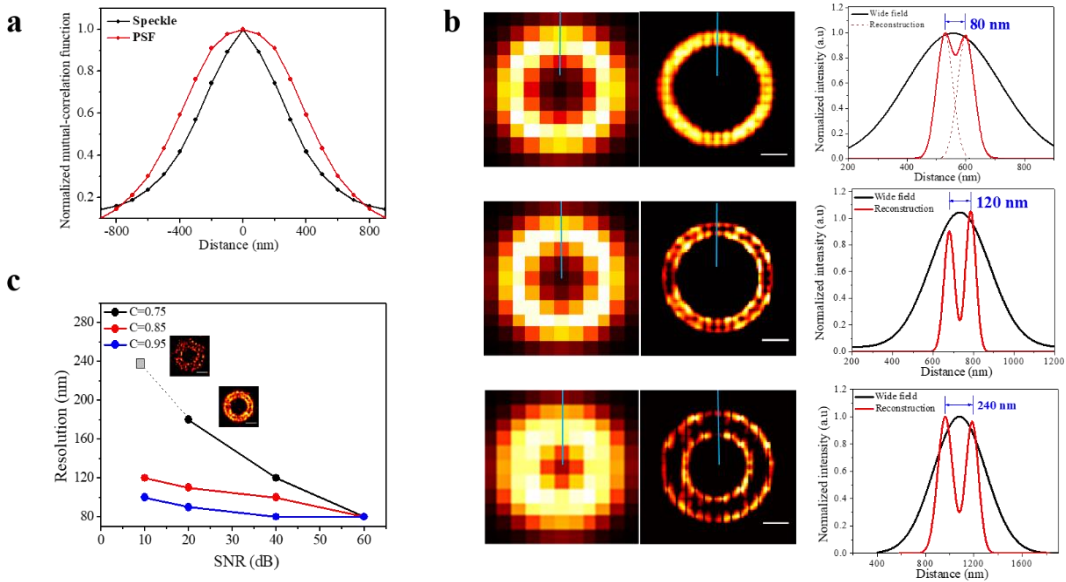

**Fig. 2.** The imaging resolution and its influencing factors for GISC nanoscope. (**a**) The normalized mutual-correlation curve comparing the performance for spatial resolution of speckle pattern and PSF. The resolution of speckle can be increased by a factor of $\sqrt{2}$ compared to PSF. (**b**) Diffraction limited rings with different spacing (80 nm, 120 nm, 240 nm) are created and single frame imaging results reconstructed by GPSR algorithm are shown in middle column, scale bar: 500 nm, corresponding to the wide field images shown in left column (160 nm/pixel), the spatial resolution of 80 nm is determined by the Rayleigh criterion shown in right column, the black and red lines respectively represent the normalized intensity distribution along the blue line in the wide field and the reconstructed images. (**c**) The effect of different *SNR* and *C* on the resolution. The insert images respectively show the reconstruction images at *SNR*=20, *C*=0.75 and *SNR*=10, *C*=0.75, the right image represents the threshold for accurate recovery using compressed sensing, the left image and the dashed lines respectively indicate the results and extents that cannot be accurate recovery using compressed sensing.

In order to clarify the factors influencing the spatial resolution of the technique, signal-to-noise ratio ($SNR$) and contrast ($C$) are investigated. $SNR = 10\log_{10}\left(\overline{I_0}/\overline{I_n}\right)$ is the main influence factor of compressed sensing algorithm, where $\overline{I_0}$ is the mean value of speckle intensity, $\overline{I_n}$ is the mean value of detected noise intensity. $C = \sigma_I / \overline{I_0}$ is the key metric to evaluate the characteristics of speckle [38], where $\sigma_I$ is the standard deviation of the speckle intensity. $C$ represents the level of the fluctuation of the speckle intensity and the $C$ of the detected speckle pattern is mainly related to the sparsity $K$ of fluorescence emitters.

To evaluate the effects of different $SNR$ and $C$ on the spatial resolution, we obtained the minimal separation distance of molecules under varying emitter densities (corresponding to different $C$) and $SNR$ by adding different levels of background noise to the speckle image. The additive noise follows Gaussian distribution, which matches the experimentally detected result by sCMOS. In **Fig. 2c**, the effect of $SNR$ on the resolution is negligible when $C$ is 0.95, equal to an emitter density of 5-8 $\mu m^{-2}$. However the $SNR$ has greatly effect on the resolution (reach to 180 nm at a $SNR$ value of 20) when $C$ is 0.75, corresponding to an emitter density above 100 $\mu m^{-2}$. The threshold for accurately reconstruction is a $SNR$ value of 20 and a $C$ value of 0.75.

Different from obtaining the continuous intensity distribution of the original signal by GPSR algorithm, we use OMP algorithm to achieve exact molecule localization expressed in discrete grid points with intensity information, and it can identify almost all completely overlapped emitters shown in **Fig. 3a**. The image resolution is determined by both the density of identified molecules (through the Nyquist sampling criteria) and the localization precision of molecules [15]. Given that our resolution is 80 nm, this corresponds to an emitter density of 156 $\mu m^{-2}$. However, in the actual simulation, the molecule identification efficiency is slightly worse due to OMP algorithm limitations and $SNR$. To quantify the ability of increasing emitter density and test the robustness of this technique against noise, we randomly generated molecules of different density in an 86×86 pixel area under different $SNR$ values, and the number of photons matches that in the experiment of Alexa-647. For each value of $SNR$, we analyze the density of identified molecules and the localization precision as the density increases. In **Fig. 3b**, the maximum density of recovered molecules is 143 $\mu m^{-2}$ at a high $SNR$ of 60, and the relatively worse densities of 76 $\mu m^{-2}$ and 62 $\mu m^{-2}$ when the $SNR$ down to 40 and 20 respectively. The localization precision at high density is still 25 nm ($SNR$=60). The lower $SNR$ leads to a worse localization precision, however at very low $SNR$ case the localization precision does not exceed 60 nm, which means that the localization precision is not the factor limiting the spatial resolution (80 nm). Therefore the ultra-high density of molecules localization is sufficient for single-frame super-resolution imaging with the clearly distinguished structures and features of the sample, and the high localization precision can maintain the imaging resolution. The ability of GISC nanoscopy in scenarios of increasing localization density and precision has orders of magnitude improvement compared to other high-density localization methods for super-resolution imaging.

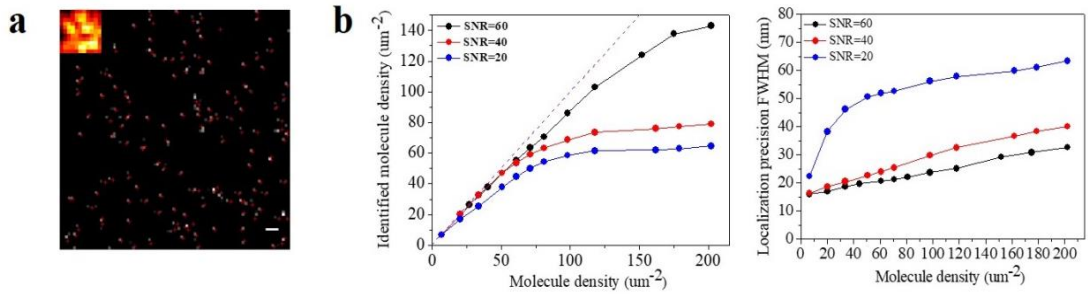

**Fig. 3.** The capability of GISC nanoscopy to identify molecules efficiently at a high density and the effect of *SNR*. (**a**) The comparison of the reconstruction result (white grids) with the true molecule positions (red crosses) at 50.7 μm⁻², scar bar: 100 nm. The smaller figures (Upper left corner) correspond to the wide field image (160 nm/pixel). (**b**) The density of identified molecules and the localization precision versus different densities and different *SNR* by OMP localization algorithm (*C*=0.75).

### 3.2 Experimental imaging

We demonstrate the performance of GISC nanoscopy on single frame super-resolution imaging of 270 nanometer ruler (GATTA-Confocal 270R, Atto 647N, Gattaquant), 160 nanometer ruler (GATTA-SIM 160R, Alexa 647, Gattaquant) and actins (labeled with stain™ 488-Phalloidin) with high labeling densities in fixed adipose-derived stem cells. Sample were imaged using a 532 nm or 640 nm laser with an intensity of approximately 0.1-0.3 kW cm⁻². 20 nm microsphere (535/575 nm or 660/680 nm) is used for calibration before imaging process. Single frame super-resolution image is reconstructed by OMP algorithm from one speckle image of sample in imaging process. **Fig. 4a** respectively shows the reconstruction results of 270 nm ruler and 160 nm ruler compared to the corresponding wide-field imaging directly recorded by EMCCD. For 270 nm ruler, we can achieve the high reconstruction probability (95%) with small localization precision (270 nm±10 nm), by comparison, the slightly lower reconstruction probability (60%) and localization precision (160 nm±30 nm) and more noisy points for 160 nm ruler. The localization precision for nanometer ruler is derived from the error of ruler and the low *SNR* due to low emitter density (corresponding to *SNR*=20, *C*=0.75). **Fig. 4b,c** show the super-resolution images of actins (labeled with stain™ 488 in Adipose-derived stem cells) are reconstructed by OMP algorithm respectively from one speckle image. Visually comparing with corresponding wide-field images GISC nanoscopy exhibits better spatial resolution and clearly separation of closely adjacent subwavelength features of actins. And the spatial resolution is up to 89 nm by fitting the intensity distribution of reconstruction molecules (along the blue line in **Fig. 4b,c**) with Gaussian peaks to obtain the full-width at half-maximum (FWHM) values as the size of actin resolved.

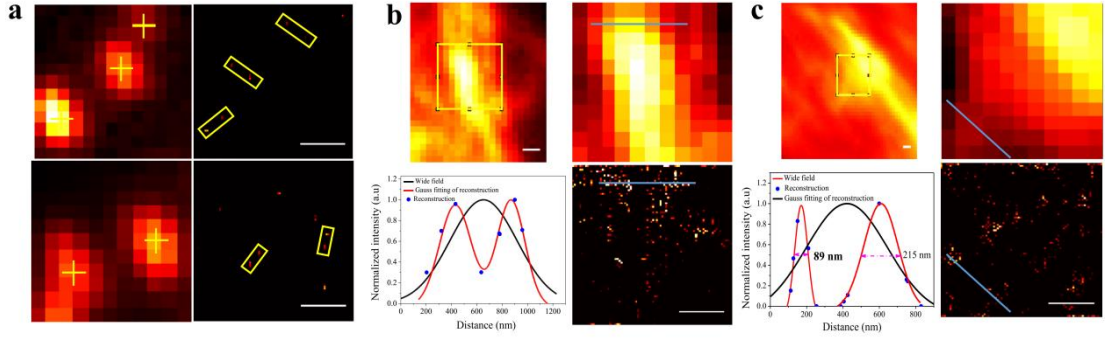

**Fig. 4.** Experimental results of nanometer rulers and actins. (**a**) The reconstruction results (Right) of 270 nanometer ruler (Upper row) and 160 nanometer ruler (Lower row) at *SNR*=20, *C*=0.75, corresponding to the wide field images (left), the yellow boxes represent the reconstruction results of diffraction limited rulers, corresponding to '+' in the wide field images. (**b**) and (**c**) show comparison between the reconstruction results (Lower right) of actins with high label densities at *SNR*=35, *C*=0.85 and the wide field images (left and upper right), the yellow boxes(left) represent the actual reconstruction areas, corresponding to the images (upper right). The intensity profiles (Lower left) comparing the images recovered from single frame and the wide field images. Normalized cross-sections taken along the solid blue lines, the black line corresponds to the normalized intensity distribution of wide field images, the blue points and the red lines respectively correspond to the normalized intensity distribution and the Gaussian fitting of the reconstruction results, in which this technique can recover the profiles of sub-diffraction spaced actins and the size of actin resolved reach to 89 nm. 160 nm per pixel in wide field images, the scale bars in other images indicate 500 nm.

### 3.3 Fast GISC-STORM

The capability of ultra-high density localization can not only achieve single frame super-resolution imaging, but also develop a fast nanoscopy combining with STORM (GISC-STORM), which utilize stochastically blinking emitters of high density to further improve the spatial resolution than single frame. The performance of GISC-STORM is characterized by the identified density and localization precision of molecules, we performed simulations to evaluate these two metrics with the molecule density ranging from 1 to 12.5 $\mu m^{-2}$ and make a fair comparison with other molecule identification methods such as CS-STORM [15] and single-molecule fitting method (implemented using ThunderSTORM [39]). As shown in **Fig. 5a**, GISC-STORM outperforms the other methods in terms of high density localization. ThunderSTORM only identifies molecules at a maximum density of 1 $\mu m^{-2}$ then follows decreasing trend under further increasing molecule density and CS-STORM also follows similar trend when its molecular density exceeds 9 $\mu m^{-2}$. However the identified molecules density of GISC-STORM has a linear increase until to the maximum density of 143 $\mu m^{-2}$ (not shown here but shown in **Fig. 3b**) which up to 16 fold compared to CS-STORM and 135 fold compared to ThunderSTORM. For localization precision, **Fig. 5b** shows that as increasing molecule density GISC-STORM maintain the localization precision lower than 20 nm while others follow continuously increasing trends (up to 98 nm precision by CS-STORM and 114 nm by ThunderSTORM). Therefore for GISC-STORM the ability of high localization density and precision implies a great reduction in the required sampling frames for the reconstruction of a super-resolution image and the spatial resolution is not limited by the localization precision at high emitter density. To further assess the ability of GISC-STORM on the improved sampling speed, we perform a qualitative comparison with CS-STORM and ThunderSTORM by respectively testing the required minimum number of frames to reconstruct a given ring with a spacing of 60 nm. The ring is composed of blinking emitters and the emitter densities per frame are determined by their respective identified molecule density (shown in the above simulation results). **Fig. 5c** shows the simulation result that a resolution of 60 nm can be achieved from 10 frames by GISC-STORM, about 50-fold faster than CS-STORM and about 400-fold faster than ThunderSTORM.

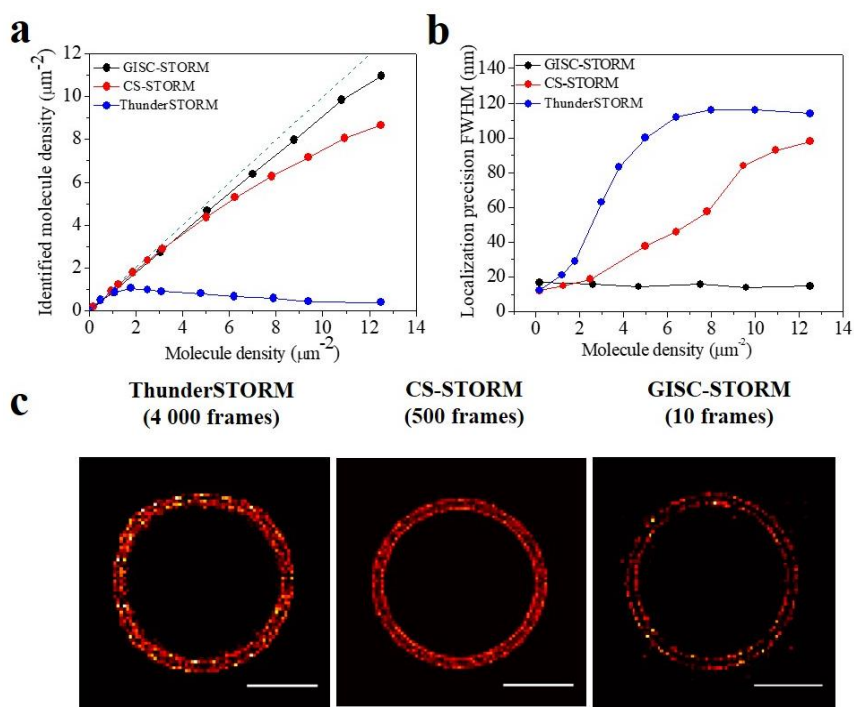

**Fig. 5.** Simulation results that demonstrate the capability of GISC-STORM to identify molecules efficiently at a high density. (**a**) Comparison the performance of GISC-STORM, CS-STORM and singe-molecule fitting method in efficiency of molecule identification. The dashed line marks the case when the number of identified molecules equals to the number of molecules. (**b**) Comparison of localization precisions. (**c**) Reconstruction results of the ring with a spacing of 60 nm from 4 000 frames by single-molecule fitting method, 500 frames by CS-STORM and 10 frames by GISC-STORM respectively. Scale bar: 500 nm.

In experiment, GISC-STORM can resolve 40 nm ruler (GATTA-PAINT 40R, Atto 655, Gattaquant) from only 100 frames in **Fig. 6a**. And **Fig. 6b** shows a qualitative comparison between GISC-STORM and ThunderSTORM for actins (labeled with Alexa 647 in U2OS cells) of high emitter density. Clearly, the sampling time of GISC-STORM to resolve the structures of actins is down to 200 frames and the localization precision of single molecule is about 20 nm (the size of one pixel), much better than ThunderSTORM image reconstructed from 200 frames and 1 000 frames respectively. Moreover, the calculation time of GISC-STORM takes about 3-4 minutes for a super-resolution image from 200 frames. Therefore GISC-STORM can significantly improve the temporal resolution (only need tens to hundreds of frames) and not come at the expense of its spatial resolution, which can promote the application of STORM in living cells imaging.

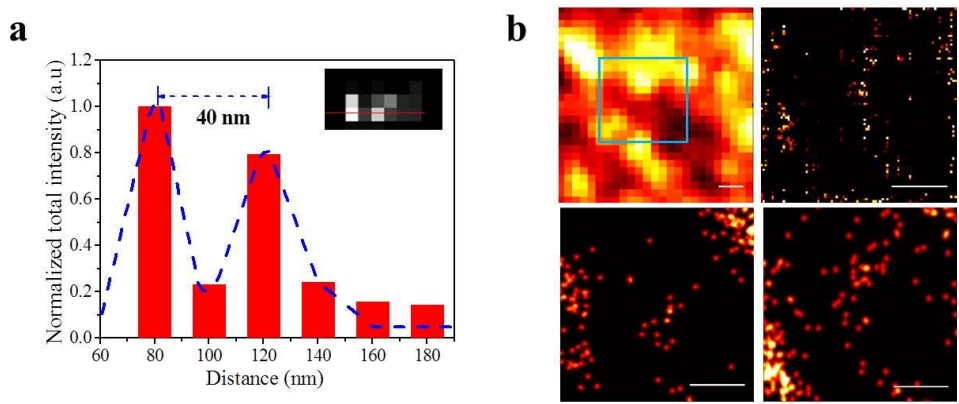

**Fig. 6.** Experimental results that demonstrate the capability of GISC-STORM to reduce the sampling time. (**a**) The reconstructed result (Upper right corner) and histogram for one representative DNA origami structures at a designed distance of 40 nm from 100 frames, the histogram shows the accumulated intensity of the red line in the insert image (20 nm/pixel). (**b**) The experimental results of actins in U2OS cells with high density labeled of Alexa 647, comparing the GISC-STORM image from 200 frames in upper right, and the STORM image using ThunderSTORM respectively from 200 frames in lower left and 1 000 frames in lower right, the raw image during STORM data acquisition in upper left, in which the blue box represents the reconstruction area. Scale bar: 500 nm.

## 4. Discussion

In this article, we developed a single frame wide-field nanoscopy based on ghost imaging via sparsity constraints by a random phase modulator to obtain speckle patterns of fluorescence signals, which can provide a resolution of 80 nm at high *SNR*. The ability to improve resolution is demonstrated on both simulated and experimental data. Moreover, we are able to reconstruct super-resolved structures with high emitter densities and low computational cost using OMP algorithm of CS, the density can reach 143 um$^{-2}$ meanwhile the localization precision is below 25 nm, which is a significant improvement compared to other super-resolution microscopes based on localization of single molecules.

GISC nanoscopy not only enables ultra-fast super-resolution imaging down to single frame, but also can obtain high resolution by combining with STORM and greatly improve STORM imaging speed, which can promote the application of super-resolution imaging in the study of living cells and microscopic dynamic processes. Moreover, this technique can be used for any fluorescent sample without variations in brightness either induced by photoactivation or intrinsic. The instruments required for imaging can easily be combined with conventional microscope.

**Funding**

## Acknowledgements

We would also like to thank Y. Liu (iHuman Institute of ShanghaiTech University), J. Li and W. Wei (Shanghai Jiao Tong University) for providing the sample, W. Ji (Institute of Biophysics, Chinese Academy of Sciences) for helping with sample and reading of the manuscript, J. Ma (Fudan University) for discussion and helpful comments.

## Disclosures

The authors declare that there are no conflicts of interest related to this article.